\documentclass[reprint,showpacs,preprintnumbers, amsmath,amssymb, aps,prb,]{revtex4-1}

\usepackage{graphicx}
\usepackage{dcolumn}
\usepackage{bm}
\usepackage{color}

\begin{document}

\title{Electron backscattering in a cavity: ballistic and coherent effects}

\author{A.~A.~Kozikov$^1$, D. Weinmann$^2$, C.~R\"{o}ssler$^1$, T.~Ihn$^1$, K.~Ensslin$^1$, C.~Reichl$^1$ and W.~Wegscheider$^1$}
\affiliation{$^1$ Solid State Physics Laboratory, ETH Z\"{u}rich, CH-8093 Z\"{u}rich, Switzerland}%
\affiliation{$^2$ Institut de Physique et Chimie des Mat\'{e}riaux de Strasbourg, Universit\'{e} de Strasbourg, CNRS UMR 7504, 23 rue du Loess, F-67034 Strasbourg, France}%


\begin{abstract}
Numerous experimental and theoretical studies have focused on low-dimensional systems locally perturbed by the biased tip of a scanning force microscope. In all cases either open or closed weakly gate-tunable nanostructures have been investigated, such as quantum point contacts, open or closed quantum dots, etc.
We study the behaviour of the conductance of a quantum point contact with a gradually forming adjacent cavity in series under the influence of a scanning gate. Here, an initially open quantum point contact system gradually turns into a closed cavity system. We observe branches and interference fringes known from quantum point contacts coexisting with irregular conductance fluctuations. Unlike the branches, the fluctuations cover the entire area of the cavity. In contrast to previous studies, we observe and investigate branches under the influence of the confining stadium potential, which is gradually built up. We find that the branches exist only in the area surrounded by cavity top gates. As the stadium shrinks, regular fringes originate from tip-induced constrictions leading to quantized conduction. In addition, we observe arc-like areas reminiscent of classical electron trajectories in a chaotic cavity. We also argue that electrons emanating from the quantum point contact spread out like a fan leaving branch-like regions of enhanced backscattering.

\end{abstract}

\maketitle

\section{\label{}INTRODUCTION}

In scanning gate \cite{Eriksson} experiments a biased tip couples capacitively to the electron gas under the sample surface and acts as a movable top gate \cite{TopinkaNat, TopinkaSci,Woodside,Pioda2004,Fallahi,Aoki,HackensNat,Pioda,Gildemeister,Bleszynski,SchnezPRB1,Pascher,GG,GG1,Pascher1,Sellier}. When negatively biased, it usually reduces the conductance of the nanostructure compared to that unperturbed by the tip (see e.g. \cite{Eriksson,Sellier,KozikovBranches}). A remarkable tip-induced enhancement of the conductance was observed in ballistic stadii \cite{KozikovStadium} presumably due to the tip-enhanced adiabaticity of the structure \cite{QPCseries}. The effect was predicted to occur when an unperturbed quantum point contact is tuned to a conductance between plateaus \cite{Jalabert2010}. The change in the conductance caused by the tip occurs already several microns away from the structure \cite{KozikovBranches} due to the ``tails'' of the tip-induced potential. When strongly biased, the tip also depletes the electron gas beneath and acts as a movable backscatterer. In the case where the tip is close to a quantum point contact (QPC), electron waves scatter off the tip-depleted disc back into the QPC and reduce the conductance. Narrow spatial regions where this effect is strong are termed branches \cite{TopinkaNat}. The interference of electron waves backscattered off the tip-depleted disc, off the QPC and impurities modulates the transmission probability of the latter and leads to the appearance of interference fringes along branches \cite{TopinkaNat, TopinkaSci,KozikovBranches, Paradiso}.

Scanning gate microscopy (SGM) studies have focused on completely open systems, such as quantum point contacts \cite{TopinkaNat, TopinkaSci, Pascher1, Sellier, KozikovBranches,Jalabert2010,Paradiso,JuraNat,JuraPRB1,JuraPRB2,Pascher2,SzafranQPC, Gorini, Heller}, and more closed systems, for example, cavities \cite{KozikovStadium, Crook, Ferry1, Ferry2,SzafranCavity,Peeters,KozikovAB}. In cavities irregular conductance fluctuations \cite{KozikovStadium, Crook, Ferry1, Ferry2} and regular fringes \cite{KozikovStadium, KozikovAB, Steinacher2015} originating from quantized tip-induced constrictions were observed.
In this paper we make a bridge between these QPC and cavity systems. The evolution of the system conductance perturbed by the biased tip is studied as we gradually form a circular cavity adjacent to a QPC at a temperature of 300 mK.
For the QPC we observe the well known branches and interference fringes, whose behavior we study as a function of the cavity confining potential. When the cavity is formed, irregular conductance fluctuations appear, which coexist with the branches and, unlike them, cover the entire area of the cavity. As the stadium is made smaller, we start observing regular fringes at its larger opening due to the formation of additional narrow constrictions between the tip-depleted region and the cavity boundaries.
In addition, we observe arc-like structures in the conductance images. They resemble the shape of classical trajectories of electrons injected into the cavity and bouncing off its boundaries. We discuss the origins of the observed effects. Although the measured scanning gate maps are complex and contain many different features, in this paper we focus on those that stand out against conductance fluctuations as we discuss in the following.

The overall effect of the biased tip on the conductance is not straightforward to interpret. In the literature the branches are often considered as regions where electrons flow in the absence of the tip (local current density) \cite{TopinkaNat, TopinkaSci, Aidala, Metalidis, Cresti, Metzger, Maryenko, See, Scannell, Micolich}.
We therefore use the obtained results to again raise the question of the interpretation of SGM maps \cite{Jalabert2010,Peeters,SzafranDot} and more specifically whether branches detected in the presence of the tip reflect electron trajectories in the absence of the tip.

\section{\label{}EXPERIMENTAL METHODS}

The structure under study is fabricated on a high-mobility GaAs/AlGaAs heterostructure using electron-beam lithography. The two-dimensional electron gas is located 120~nm below the surface and has a mobility of $8\times10^6$ cm$^2$/Vs at an electron density of $n = 1.2\times10^{11}$ cm$^{-2}$ at 300 mK. The corresponding elastic mean free path and the Fermi wavelength are $l_\mathrm{p}=49~\mu$m and $\lambda_\mathrm{F}=72$ nm, respectively.

\begin{figure}[t]
\begin{center}
\includegraphics[width=5cm]{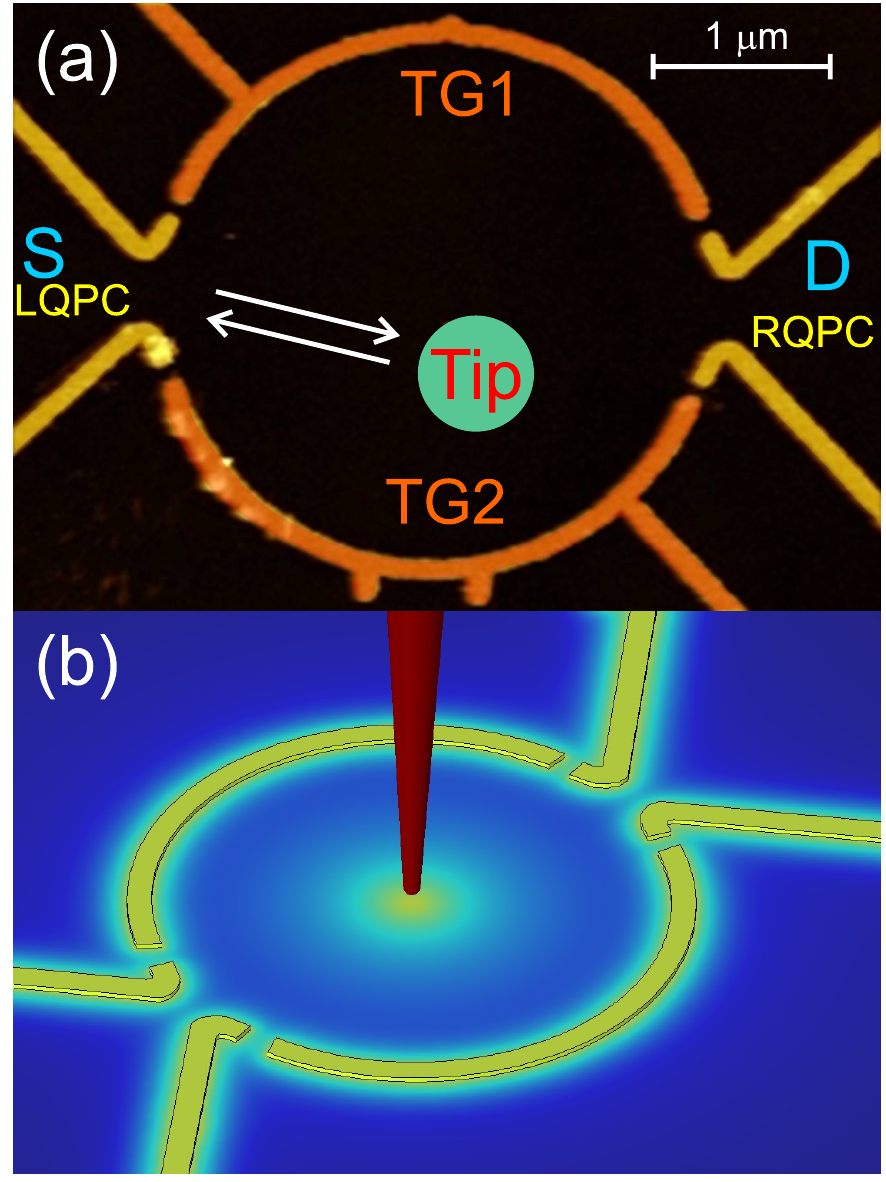}
\caption{(a) Room temperature AFM image of the studied sample. Pairs of top gates form the left QPC (LQPC), right QPC (RQPC) and stadium (TG1 and TG2) and are marked by different colors. The tip-induced depletion region (bright green) backscatters electrons injected into the stadium from the left QPC (white arrows). The ac current flows between the source ``S'' and drain ``D'' contacts. (b) Schematics of an SGM setup. A conical tip (dark red) is placed above the surface (blue) inside the stadium (yellow metallic top gates). The negatively biased tip produces a strong perturbation in the 2DEG. Below the tip the 2DEG is completely depleted (a circular region below the tip). Such a disc acts as a moveable backscatterer. The long-range nature of the tip-induced potential is indicated as a color gradient between the tip-depleted region and the top gates. The greenish area around the top gates marks the depletion around them.}
\label{fig:Sample}
\end{center}
\end{figure}

Experiments are performed on a circular cavity defined by surface gates TG1 and TG2, and two narrow constrictions (LQPC and RQPC) at its entrance and exit as shown in Fig. \ref{fig:Sample}(a). The lithographic diameter of the cavity and the width of the constrictions are $D=3~\mu$m and $W=300$ nm, respectively, much smaller than $l_\mathrm{p}$.
The electronic structures are defined electrostatically by applying a negative voltage to the top gates which are 30 nm high and 150 nm wide.

Scanning gate experiments are carried out at a temperature of 300 mK in a He-3 refrigerator using a home-built scanning force microscope \cite{Ihn}. The conductance across the sample $G$ is measured using standard lock-in techniques in a two-terminal configuration by applying an AC rms source-drain voltage of 100 $\mu$V at a frequency of about 27 Hz. The metallic tip is placed 60 nm above the GaAs surface and biased to -8 V [Fig. \ref{fig:Sample}(b)]. At this voltage the induced potential at the Fermi energy is steep enough to observe sharp interference fringes and branches. The electron gas below the tip is depleted resulting in a depletion disc of about 0.5 $\mu$m in diameter \cite{Steinacher2015}, which is smaller than the size of the cavity $D$. The situation when the tip-depleted region and the cavity have comparable sizes was thoroughly studied in a different sample in our previous works \cite{KozikovAB, Steinacher2015}. The tip scans the surface at a constant height and the conductance is recorded simultaneously producing 2D maps $G(x,y)$ as a function of tip position $(x,y)$. The contact resistance has been subtracted in all figures.

\begin{figure*}[t]
\begin{center}
\includegraphics[width=\textwidth]{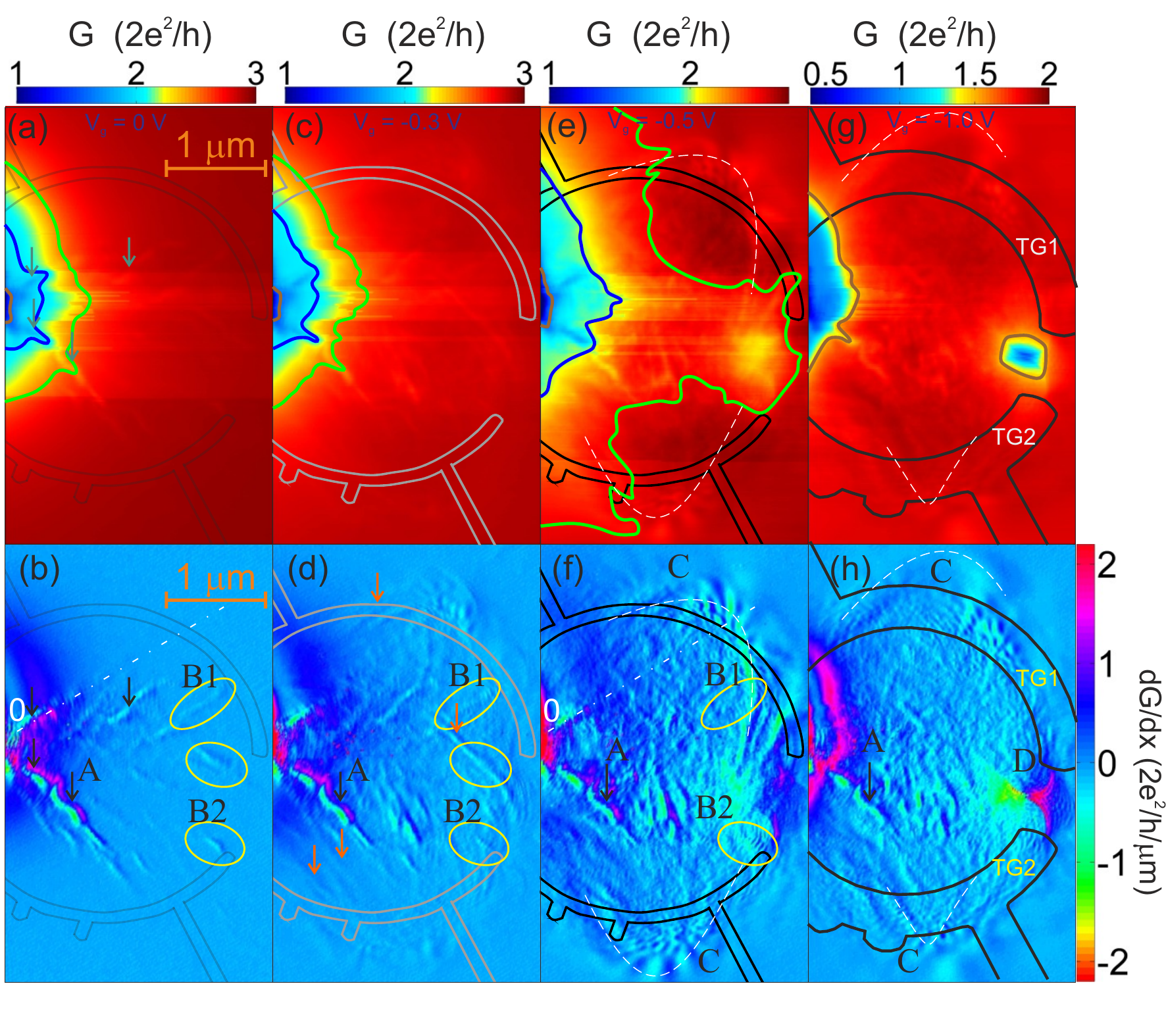}
\caption{Evolution of the QPC conductance $G(x,y)$ (upper panel) and its numerical derivative $\mathrm{d}G(x, y)/\mathrm{d}x$ (lower panel) with respect to the scan direction $x$ as the voltage applied to the stadium $V_\mathrm{g}$ varies from 0 to -1 V. Arrows in (a) and the lower panel and yellow ellipses in the lower panel mark branches (fringes) of type A and B, respectively. Dash-dotted lines in (b) and (f) indicate regions where the 1D conductance is studied quantitatively. The label ``0" near one end of this line marks the beginning of the 1D cuts. Curved dashed lines in (e)-(h) mark the position of the arc-like areas of type C. The QPC is biased to the 3rd conductance plateau without the tip and the stadium. Biased top gates are shown as solid lines. The thickness of the top gates in (g) and (h) is enhanced to illustrate the depletion zone for negative gate voltages. Brown, green and blue contour lines in the top panel correspond to the constant conductance of 1.5, 2 and 2.5$\times$2$e^2$/h. Green contour lines in (e) in the arc-like areas are intentionally absent to avoid confusion.}
\label{fig:QPCstadium}
\end{center}
\end{figure*}

Figure \ref{fig:QPCstadium} shows the dependence of the conductance $G$ as the voltage $V_\mathrm{g}$ applied to TG1 and TG2 is varied between 0 and -1 V. The left constriction, LQPC, is set to the third conductance plateau (without the stadium and the tip). The gates of the right constriction, RQPC, are kept grounded. When $V_\mathrm{g}$ is zero, conductance measurements through the LQPC in Fig. \ref{fig:QPCstadium}(a) show a set of branches (some of them are marked by green arrows) on top of a slowly varying background (see contour lines). This background originates from the ``tails'' of the tip-induced Lorentzian potential capacitively coupled to the QPC (gating effect). The gating effect reduces $G$ from about 3 to 1.3$\times2e^2/h$ (when the tip approaches the constriction LQPC) on a length scale of microns. The branches are a consequence of electron backscattering off the tip-depleted region, which causes a drop of the conductance of the order of $0.1\times2e^2/h$ on a length scale of 100 nm. To enhance the visibility of the branches, the conductance is numerically differentiated with respect to the scan direction $x$ [Fig. \ref{fig:QPCstadium}(b)]. Due to charge rearrangements seen as horizontal lines of small conductance discontinuities in the top panel of Fig. \ref{fig:QPCstadium}, we do not differentiate in the $y$-direction. This procedure also reveals interference fringes decorating the branches, i.e. small variations of the conductance (on the order of $0.01\times2e^2/h$) on a length scale of $\lambda_\mathrm{F}/2$. We will refer to these features, which are located close to the LQPC, as fringes and branches of type A. They are well-known in scanning gate experiments on QPCs \cite{TopinkaNat, TopinkaSci, KozikovBranches,Paradiso,JuraNat,JuraPRB1,JuraPRB2} and related to electron backscattering. Fringes farthest away from the LQPC [encircled in (b)] located closer to the right-hand side opening of the cavity will be referred to as type B.

It is remarkable that although $V_\mathrm{g}=0$ V the branches only exist in the area surrounded by the cavity gates for all parameter regimes investigated (different tips, tip voltages, tip-surface distances and thermal cycles). This could be due to a shallow potential barrier below the cavity gates due to the presence of metallic electrodes or strain fields \cite{SteinacherPRB}.
The tip-depleted region is larger than the width of TG1 and TG2 and cannot be fully screened by the grounded gates. In our previous studies \cite{KozikovBranches} on LQPC branches were seen more than 3 $\mu$m away from the QPC when probed to the left of the constrictions where there is no such confinement. Interestingly, the branches follow two preferential directions. They seem to be oriented parallel to the LQPC top gates (see Fig. \ref{fig:Sample}).

When a voltage $V_\mathrm{g}$ = -0.3 V is applied to the gates TG1 and TG2 (the electron gas below the gates is depleted at -0.4 V), the background conductance decreases due to the cross coupling between the cavity and QPC top gates. Local variations of $G$ become stronger as seen by the colorcode in Fig. \ref{fig:QPCstadium}(c). These variations are more clearly seen in $\mathrm{d}G(x, y)/\mathrm{d}x$ in Fig. \ref{fig:QPCstadium}(d) as conductance fluctuations between the branches. In some areas new branches have appeared compared to Fig. \ref{fig:QPCstadium}(a) [marked by orange arrows (d)]. They are not decorated with interference fringes, but still follow the same two preferential directions. Fringes (branches) of type A look similar to those at $V_\mathrm{g}$ = 0 V, whereas fringes of type B are more pronounced.

At more negative gate voltages, $V_\mathrm{g}$ = -0.5 V, the cavity with a diameter of $D\approx2.9~\mu$m forms [Fig. \ref{fig:QPCstadium}(e)]. Together with the further decrease of the background conductance, at the right opening of the stadium $G(x,y)$ drops to about 2.1$\times2e^2/h$ due to partially closing this constriction by the tip-depleted region. The absence of this drop in Fig. 2(c) indicates that the current flow is still quite homogeneous below the gates and through the right constriction, in which it does not concentrate strongly. At $V_\mathrm{g}$ = -0.5 V when the tip is in close proximity to gates TG1 and TG2, additional arc-like areas appear (delineated by white curved dashed lines in (e) as guides to the eye) with a drop of $G$ of about $0.1\times2e^2/h$. Along and across them the conductance is modulated (fringes). We will refer to these areas and fringe-like patterns as those of type C and discuss them phenomenologically. In some regions they are outside the cavity due to the broad tip-induced potential. The arc-like areas are already seen for $V_\mathrm{g}$ = -0.3 V when the cavity has not been formed yet. In addition, they exist when the number of the QPC modes is tuned to 2, 3 and 4. Two dark spots between the arc-like areas and the center of the cavity (symmetric regions of the high conductance delineated by white dashed lines and solid green lines in (e) and (f)) at $V_\mathrm{g}$ = -0.5 V was previously attributed to the tip-induced adiabaticity \cite{KozikovStadium}.
In $\mathrm{d}G(x, y)/\mathrm{d}x$ in Fig. \ref{fig:QPCstadium}(f) fringes (branches) of type A look unchanged and the amplitude of those of type B continue increasing in the encircled region B2. The number and amplitude of conductance fluctuations grow, which mask some of the fringes, e.g. those in the encircled region B1. These fringes are still present, but more difficult to see. Though electrons in the cavity are ballistic, conductance fluctuations arise from the interference of electrons bouncing off the cavity boundaries. Another drastic difference between $V_\mathrm{g}$ = -0.3 and -0.5 V is that branches in the latter case do not have a clear direction apart from those close to the LQPC. Branches B1 and B2 seem to be an exception. The classical dwell time of electrons in the cavity region increases drastically as the cavity forms. Trajectories bouncing off the cavity boundaries several times become more important. In presence of the tip or weak disorder, the classical dynamics therefore becomes more chaotic.

As the cavity is made smaller ($D\approx2.7~\mu m$) by applying a more negative gate voltage of $V_\mathrm{g}$ = -1.0 V [Fig. \ref{fig:QPCstadium}(g) and (h)], the number of transmitted modes in the LQPC that contribute to transport continues to become smaller. Previously \cite{KozikovBranches, KozikovNewTechnique}, we showed a QPC in the absence of an adjacent cavity becoming narrower under the application of the constriction-forming gate voltage. Here we observe the same effect in the presence of an additional confining potential, the cavity. Since the number of modes is directly related to the angular range of the branches \cite{TopinkaSci}, the number of branches decreases with $V_\mathrm{g}$. Nevertheless, the remaining branches of type A survive and look the same. At the right opening of the cavity a fringe pattern appears [labelled as D in Fig. \ref{fig:QPCstadium}(h)] and the conductance there can now be suppressed indicating that the current density is channeled through this opening. As discussed in detail in Ref. \cite{KozikovStadium}, inside this pattern the tip forms two short and narrow parallel channels with TG1 and TG2. The conductance of each of them is quantized producing a single set of fringes. When transport occurs through both of these channels, two sets of fringes cross each other forming a checkerboard pattern (for details see Ref. \cite{KozikovStadium}). All the discussed observations remain if we use the right QPC (RQPC) and keep the LQPC grounded.

\begin{figure}[t]
\begin{center}
\includegraphics[width=\columnwidth]{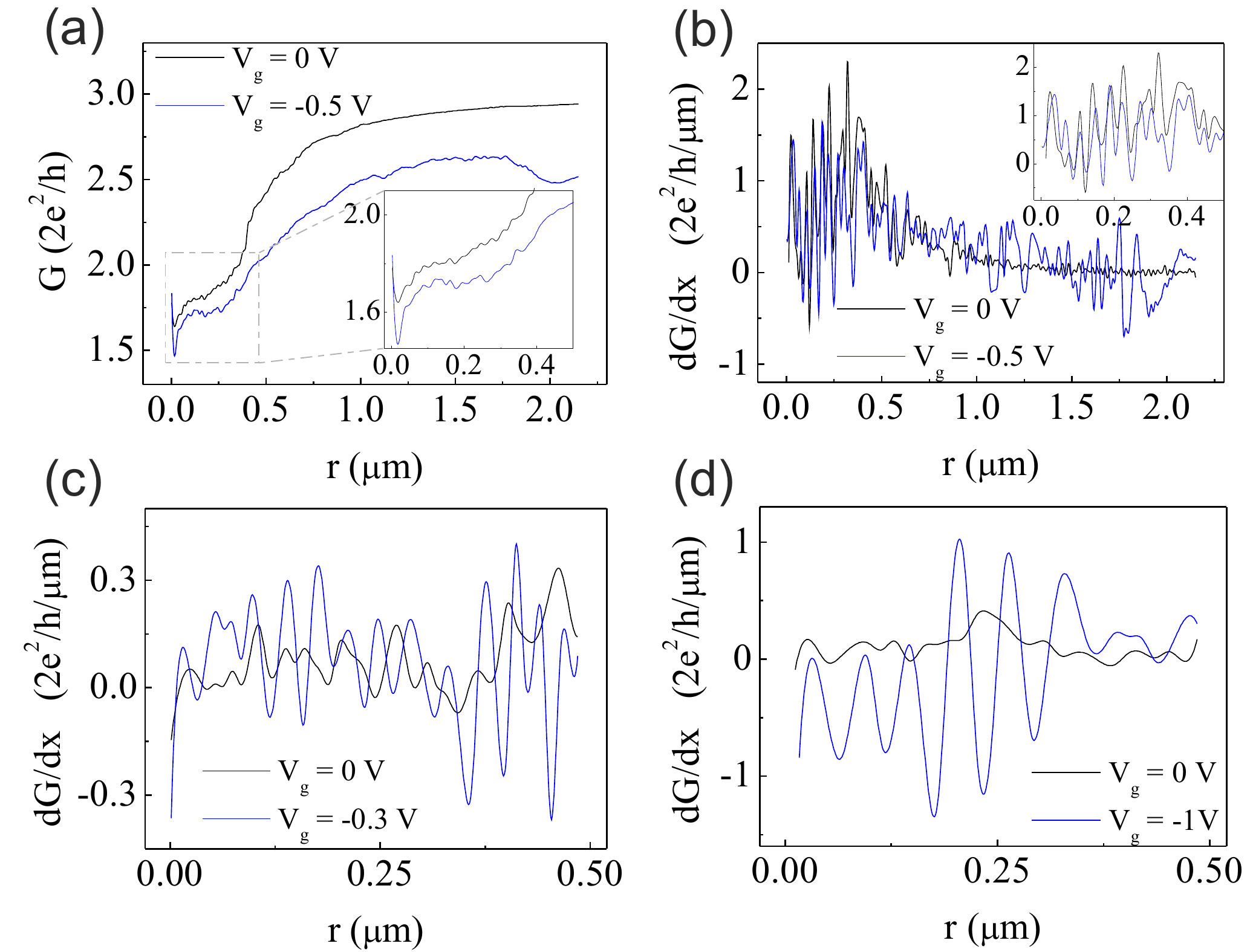}
\caption{(a) $G$ as a function of distance, $r$, along the white dash-dotted line shown in Fig. \ref{fig:QPCstadium}(b) and (f) for different $V_\mathrm{g}$. $G$ in a smaller range of $r$ (zoom-in) is given in the inset. Interference fringes at $r<0.8$ $\mu$m and conductance fluctuations at $r>0.8$ $\mu$m are visible when the stadium is formed (blue curve). (b) $\mathrm{d}G(x, y)/\mathrm{d}x$ along the white dash-dotted line shown in Fig. \ref{fig:QPCstadium}(b) and (f). Interference fringes and conductance fluctuations are more clearly visible at small and large $r$, respectively. (c) $\mathrm{d}G(x, y)/\mathrm{d}x$ across interference fringes in region B1 indicated in Fig. \ref{fig:QPCstadium}(b) and (d) at different $V_\mathrm{g}$. (d) Same as (c), but in region B2 [Fig. \ref{fig:QPCstadium}(b) and (f)].}
\label{fig:Cuts}
\end{center}
\end{figure}

The interference fringes of type A do not seem to change when the cavity forms, i.e. when the electron dynamics becomes chaotic. To check this observation,
we plot in Fig. \ref{fig:Cuts}(a) and (b) $G(r)$ and $\mathrm{d}G(r)/\mathrm{d}r$, respectively, along the white dash-dotted line [shown in Fig. \ref{fig:QPCstadium}(b) and (f)]. When the tip is close to the LQPC ($r\lesssim0.8$ $\mu$m), the amplitude and position of the interference fringes (better seen in the insets) almost do not change as the voltage applied to the stadium is varied between 0 and -0.5 V. These fringes originate from interference that involves electron waves directly backscattered off the tip-depleted region. The presence of additional barriers outside the electron paths does not affect the interference conditions.
At $r\gtrsim0.8$ $\mu$m conductance fluctuations appear, which are clearly seen in Fig. \ref{fig:Cuts}(b) (blue curve). The amplitude of these fluctuations is similar to that of the interference fringes. This is why the interference fringes very well seen at $V_\mathrm{g}=0$ V may become hidden or disappear as the voltage $V_\mathrm{g}$ is made more negative. The behaviour of the branches of type A is similar to that in a situation when the cavity gates are grounded and the LQPC becomes narrower by applying a voltage to its gates. Therefore, the right-hand side constriction does not influence this behaviour and the effect of the voltage applied to TG1 and TG2 is similar to that applied to the LQPC gates. The behavior of the interference fringes of type B is different. Their amplitude increases as the cavity shrinks. Plots of $\mathrm{d}G/\mathrm{d}x$ in Fig. \ref{fig:Cuts} (c) and (d) along cuts through regions B1 and B2 shown in Fig. \ref{fig:QPCstadium} demonstrate this observation.

We note that the conductance fluctuations and variations in the amplitude of the fringes are caused by the potential of the cavity gates despite the fact this potential changes the number of transmitted modes in the LQPC. Neither of these features was observed in our previous studies \cite{KozikovBranches, KozikovNewTechnique} in open geometries even when we compensated the cross-coupling effect of the tip on a QPC. The tip has a much stronger effect on the QPC potential than the cavity gates.

Previous theoretical calculations on open 2DEGs with a shallow random potential variation much smaller than the Fermi-energy indicated that the experimentally observed branches in scanning gate images correspond to the local current density and therefore mark regions where electrons flow \cite{TopinkaNat, Heller2}. Images obtained from our cavity structures cannot be interpreted in the same intuitive way due to multiple reflections of electrons from the cavity walls.

The branches/fringes of type B depend on $V_\mathrm{g}$, which indicates that they must involve scattering from the gate-induced cavity boundary. As a result, one would expect to see continuous branches and fringes all the way from the QPC to the boundaries of the cavity. In addition, they should spread out like a fan to account for the finite length of the type B branches. However, this behaviour is not observed. Therefore, branches and fringes occur in regions where backscattering is strong.
In addition, as the conductance fluctuations observed when the cavity forms at $V_\mathrm{g}<-0.4$~V exist in the entire area of the stadium, electrons are likely to spread out like a fan when they enter the cavity.

The coexistence of branches, interference fringes and conductance fluctuations supports the idea that the concept of branched flow of electrons becomes increasingly irrelevant with increasing confinement in our structure. It is true that for quantum point contacts, which are more open systems, the pattern in backscattering has a close relation to electron flow. In the present situation of a confined cavity, however, branches observed in backscattering not always relate to electron flow, as we discuss in detail.

\begin{figure}[t]
\begin{center}
\includegraphics[width=\columnwidth]{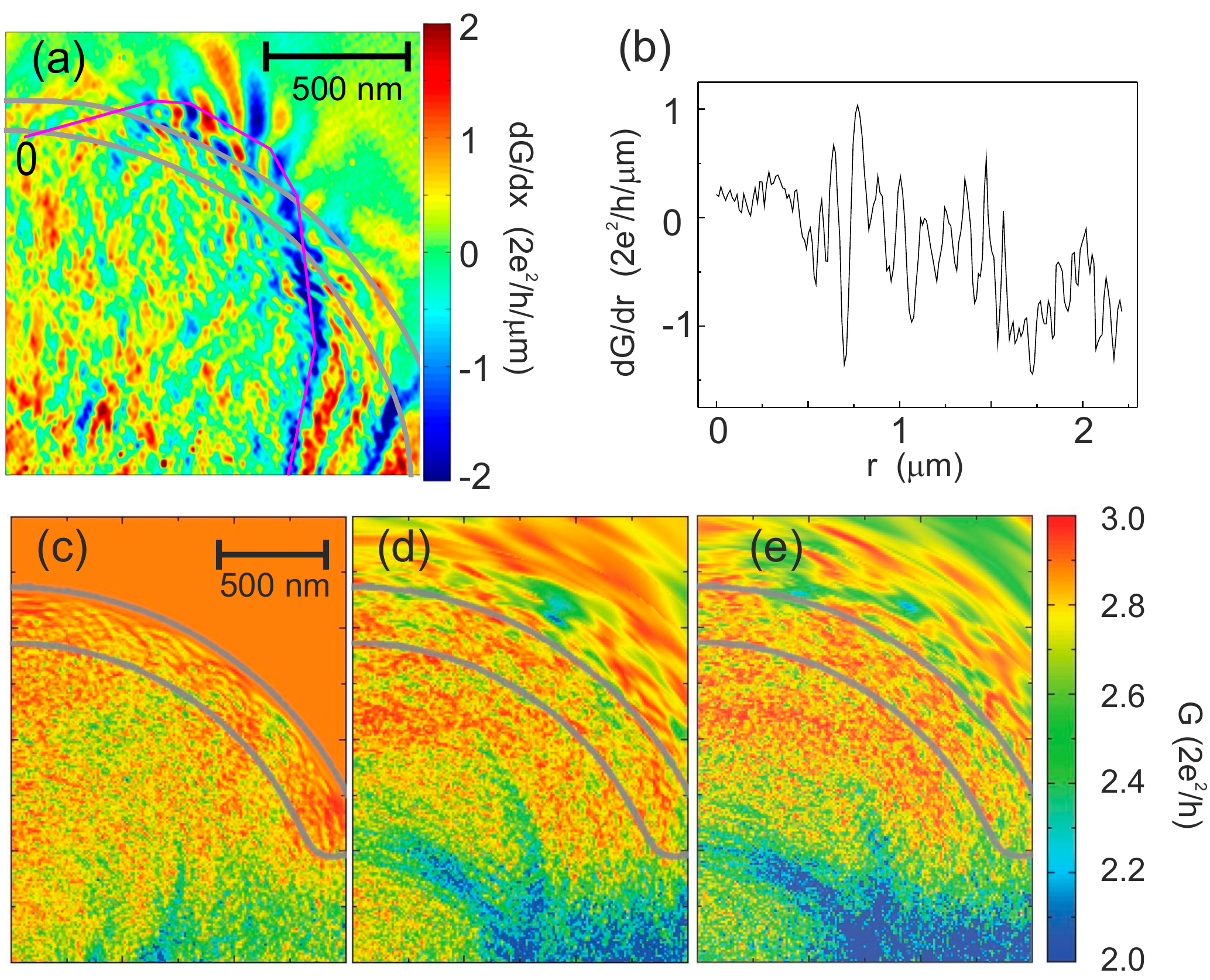}
\caption{(a) A close-up of the upper curved branch in Fig. \ref{fig:QPCstadium}(f): numerical derivative of the conductance as a function of tip position. (b) 1D cut along the pink polyline in (a). Zero in the plotted curve is marked by the label ``0" in (a). (c)-(e) Simulations of the conductance as a function of tip position for a hard-wall (c) and Lorentzian tip-induced potential (d), and including disorder (e). Grey solid lines outline the top gate TG1.}
\label{fig:SideBranch}
\end{center}
\end{figure}

The arc-like areas and fringes of type C in Fig. \ref{fig:QPCstadium}(e)-(h) are reminiscent of classical electron trajectories in a ballistic cavity: electrons entering the stadium bounce off its boundaries towards its exit. In Fig. \ref{fig:SideBranch}(a) we show a close-up of the upper arc-like area together with a 1D cut along it [Fig. \ref{fig:SideBranch}(b)]. Fringes that decorate the arc are perpendicular to its direction. But unlike type A and B, fringes of type C have a curvy and discontinuous shape. They are spaced by roughly half the Fermi wavelength (with a variation of 30$\%$ \cite{KozikovBranches}) when the tip is inside the stadium, i.e. at $r<0.5$ $\mu$m [see Fig. \ref{fig:SideBranch}(b)]. But when the tip is above the top gates, $0.5<r<1.5$ $\mu$m in (b), the fringes are not periodic and spaced by 100-200 nm. In addition, the areas of type C do not look like branches of type A or B, which are caused by electron backscattering and usually look roughly like straight lines originating from a point (QPC). We thus conclude that the observed effects of type C must have a different origin than those of type A and B.

To better understand the modelling of the scanning gate experiments, we performed
fully coherent numerical simulations within a tight-binding model using the Kwant quantum transport package \cite{Groth}. The calculated zero-temperature conductances for different shapes of the tip-induced potential and strength of disorder are shown in Fig. \ref{fig:SideBranch}(c)-(e). In Fig. \ref{fig:SideBranch}(c) we consider an idealized hard wall tip-induced potential and in Fig. \ref{fig:SideBranch}(d, e) a more realistic smooth potential of Lorentzian shape. The potentials entering this calculation induced by the top gates, and the one by the tip in (d, e), are results of a separate COMSOL finite-elements electrostatics simulation. The parameters were chosen to correspond to a LQPC conductance of $3 \times 2e^2/h$ with grounded gates TG1 and TG2 and without the tip, and $G = 2 \times 2e^2/h$ with the tip at the right opening of the cavity to mimic experimental conditions that correspond to Figs. \ref{fig:QPCstadium}(e)-(f). While a perfectly clean device is assumed in Figs. \ref{fig:SideBranch}(c, d) we present in Fig. \ref{fig:SideBranch}(e) results for the situation of (d) affected by an additional disorder potential. To model disorder, we used an Anderson model on the square tight-binding lattice having a lattice spacing of 5 nm, with independent random on-site energies that are drawn from a uniform distribution within $[-W/2,W/2]$. In (c) and (d) the disorder strength $W = 0$, whereas in (e), $W$ is chosen such that it leads to the experimental mean free path.
In all these scenarios the simulations reveal conductance fluctuations, which are qualitatively similar to those observed in the experiment, when the tip is inside the cavity. Quantitatively the simulated fluctuations have a much larger amplitude (on the order of 10 times larger) and are more abrupt possibly because the simulations do not take into account finite temperature, decoherence, electron-electron interaction and correlated disorder. The situation changes drastically for tip positions outside the cavity. The conductance $G=2.9\times2e^2/h$ remains unperturbed for the hard-wall tip-induced potential (Fig. \ref{fig:SideBranch}(c)). It is smaller than $3\times2e^2/h$ due to the biased gates TG1 and TG2. For the realistic tip potential assumed in Fig. \ref{fig:SideBranch}(d) $G$ varies between 2 and $3\times2e^2/h$ due to the long-range ``tails" of the Lorentzian as seen in Fig. \ref{fig:SideBranch}(d). Including uncorrelated disorder smears out conductance fluctuations and reduces their amplitude (Fig. \ref{fig:SideBranch}(e)). In all scenarios (c)-(e) there are positions (red regions) at which the tip enhances the unperturbed conductance of $2.9\times2e^2/h$ to almost $3\times2e^2/h$ (the conductance of the LQPC) possibly due to the tip-induced adiabaticity already observed by us in previous measurements \cite{KozikovStadium, KozikovAB}.
In the simulations the conductance varies on a large scale when the tip is outside the stadium, which is in contrast with short scale fluctuations when the tip is inside. This is vaguely similar to our experiment in which $G$ fluctuations along arc-like areas of type C outside the cavity occur on a larger scale than those inside it. The absence of the precise structure of the type C features as well as those of type A and B  in the simulations points to the difficulties of modelling the experimental situation and a possible necessity to include correlated disorder.

In conclusion, we have measured the conductance of a QPC with an adjacent cavity strongly perturbed by the biased tip. We have observed well known branches and interference fringes originating from electron backscattering off the tip-depleted region. The branches mark regions in space where backscattering dominates. They were found to have a preferential direction parallel to the QPC gates, to be restricted by the cavity top gates irrespective of the gate voltage and to coexist with conductance fluctuations inside the cavity. Unlike the branches, the conductance fluctuations cover the entire area of the stadium and depend on the stadium size. Our observations indicate that branches not necessarily reflect electron trajectories. They also point towards the fact that electrons leaving a QPC in fact spread out like a fan instead of moving only along the branches. We have also observed arc-like areas, which are reminiscent of classical electron trajectories in a chaotic cavity.
Coherent simulations carried out for different shapes of the potential induced by the tip as well as the level of disorder reproduce qualitatively conductance fluctuations inside the cavity. The origin of the arc-like areas and the corresponding fringes remain to be understood in more detail. Including correlated disorder could possibly shed more light on the nature of the observed features. Our work shows the potential of scanning gate microscopy to study the electrons behaviour in cavities. For example, it can be used to probe spin-coherent phenomena like for example micrometre-extended singlet states \cite{RoesslerPRL}.

We are grateful for fruitful discussions with Rodolfo Jalabert, Cosimo Gorini and Guillaume Weick. We acknowledge financial support from the Swiss National Science Foundation and NCCR ``Quantum Science and Technology" and the French National Research Agency ANR (Projects ANR-11-LABX-0058$\_$NIE, ANR-14-CE36-0070-01).


\begin{thebibliography}{}

\bibitem{Eriksson} M. A. Eriksson, R. G. Beck, M. Topinka, J. A. Katine, R. M. Westervelt, K. L. Campman and A. C. Gossard, Appl. Phys. Lett. {\bf 69}, 671 (1996 )
\bibitem{TopinkaNat} M. A. Topinka, B. J. LeRoy, R. M. Westervelt, S. E. J. Shaw, R. Fleischmann, E. J. Heller, K. D. Maranowski and A. C. Gossard, Nature {\bf 410}, 183 (2001)
\bibitem{TopinkaSci} M. A. Topinka, B. J. LeRoy, S. E. J. Shaw, E. J. Heller, R. M. Westervelt, K. D. Maranowski and A. C. Gossard, Science, {\bf 289}, 2323 (2000)
\bibitem{Woodside} M. T. Woodside, P. L. McEuen, Science {\bf 296}, 1098 (2002)
\bibitem{Pioda2004} A. Pioda, S. Kicin, T. Ihn, M. Sigrist, A. Fuhrer, K. Ensslin, A. Weichselbaum, S. E. Ulloa, M. Reinwald and W. Wegscheider, Phys. Rev. Lett. {\bf 93}, 216801 (2004)
\bibitem{Fallahi} P. Fallahi, A. C. Bleszynski, R. M. Westervelt, J. Huang, J. D. Walls, E. J. Heller, M. Hanson and A. C. Gossard, Nano Lett. {\bf 5}, 223 (2005)
\bibitem{Aoki} N. Aoki, A. Burke, C. R. da Cunha, R. Akis, D. K. Ferry and Y. Ochiai, J. Phys.: Conf. Ser. {\bf 38}, 79 (2006)
\bibitem{HackensNat} B. Hackens \textit{et al.}, Nature Physics {\bf 2}, 826 (2006)
\bibitem{Pioda} A. Pioda, S. Kicin, D. Brunner, T. Ihn, M. Sigrist, K. Ensslin, M. Reinwald and W. Wegscheider, Phys. Rev. B {\bf 75}, 045433 (2007)
\bibitem{Gildemeister} A. E. Gildemeister, T. Ihn, R. Schleser, K. Ensslin, D. C. Driscol and A. C. Gossard, J. Appl. Phys. {\bf 102}, 083703 (2007)
\bibitem{Bleszynski} A. C. Bleszynski, F. A. Zwanenburg, R. M. Westervelt, A. L. Roest, E. P. A. M. Bakkers and L. P. Kouwenhoven, Nano Lett. {\bf 7}, 2559 (2007)
\bibitem{SchnezPRB1} S. Schnez, J. G\"{u}ttinger, M. Huefner, C. Stampfer, K. Ensslin and T. Ihn, Phys. Rev. B {\bf 82}, 165445 (2010)
\bibitem{Pascher} N. Pascher, D. Bischoff, T. Ihn and K. Ensslin, Appl. Phys. Lett. {\bf 101}, 063101 (2012)
\bibitem{GG} A. G. F. Garcia, M. K\"{o}nig, D. Goldhaber-Gordon, K. Todd, Phys. Rev. B {\bf 87}, 085446 (2013)
\bibitem{GG1} M. K\"{o}nig, M. Baenninger, A. G. F. Garcia, N. Harjee, B. L. Pruitt, C. Ames, P. Leubner, C. Br\"{u}ne, H. Buhmann, L. W. Molenkamp and D. Goldhaber-Gordon,  Phys. Rev. X {\bf 3}, 021003 (2013)
\bibitem{Pascher1} N. Pascher, C. R\"{o}ssler, T. Ihn, K. Ensslin, C. Reichl and W. Wegscheider, Phys. Rev. X {\bf 4}, 011014 (2014)
\bibitem{Sellier} B. Brun {\it et al}, Nat. Commun. {\bf 5} 4290 (2014)
\bibitem{KozikovBranches} A. A. Kozikov, C. R\"{o}ssler, T. Ihn, K. Ensslin, C. Reichl and W. Wegscheider, New J. Phys. {\bf 15}, 013056 (2013)
\bibitem{KozikovStadium} A. A. Kozikov, D. Weinmann, C. R\"{o}ssler, T. Ihn, K. Ensslin, C. Reichl and W. Wegscheider, New J. Phys. {\bf 15}, 083005 (2013)
\bibitem{QPCseries} L. P. Kouwenhoven, B. J. van Wees, W. Kool, C. J. P. M. Harmans, A. A. M. Staring, C. T. Foxon, Phys. Rev. B {\bf 40}, 8083 (1989)
\bibitem{Jalabert2010} R. A. Jalabert, W. Szewc, S. Tomsovic and D. Weinmann, Phys. Rev. Lett. {\bf 105}, 166802 (2010)
\bibitem{Paradiso} N. Paradiso, S. Heun, S. Roddaro, L. N. Pfeiffer, K. W. West, L. Sorba, G. Biasiol, F. Beltram, Physica E {\bf 42}, 1038 (2010)
\bibitem{JuraNat} M. P. Jura, M. A. Topinka, L. Urban, A. Yazdani, H. Shtrikman, L. N. Pfeiffer, K. W. West and D. Goldhaber-Gordon, Nature Physics {\bf 3}, 841 (2007)
\bibitem{JuraPRB1} M. P. Jura, M. A. Topinka, M. Grobis, L. N. Pfeiffer, K. W. West and D. Goldhaber-Gordon, Phys. Rev. B {\bf 80}, 041303 (2009)
\bibitem{JuraPRB2} M. P. Jura, M. Grobis, M. A. Topinka, L. N. Pfeiffer, K. W. West, and D. Goldhaber-Gordon, Phys. Rev. B {\bf 82}, 155328 (2010)
\bibitem{Pascher2} N. Pascher, F. Timpu, C. R\"{o}ssler, T. Ihn, K. Ensslin, C. Reichl and W. Wegscheider, Phys. Rev. B {\bf 89}, 245408 (2014)
\bibitem{SzafranQPC} M. P. Nowak, K. Kolasi\'{n}ski and B. Szafran, Phys. Rev. B. {\bf 90}, 035301 (2014)
\bibitem{Gorini} C. Gorini, R. A. Jalabert, W. Szewc, S. Tomsovic and D. Weinmann, Phys. Rev. B. {\bf 88}, 035406 (2013)
\bibitem{Heller} B. Liu and E. J. Heller, Phys. Rev. Lett. {\bf 111}, 236804 (2013)
\bibitem{Crook} R. Crook, C. G. Smith, A. C. Graham, I. Farrer, H. E. Beere, and D. A. Ritchie, Phys. Rev. Lett. {\bf 91}, 246803 (2003)
\bibitem{Ferry1} A. M. Burke, R. Akis, T. E. Day, G. Speyer, D. K. Ferry and B. R. Bennett, Phys. Rev. Lett. {\bf 104}, 176801 (2010)
\bibitem{Ferry2} N. Aoki, R. Brunner, A. M. Burke, R. Akis, R. Meisels, D. K. Ferry and Y. Ochiai, Phys. Rev. Lett. {\bf 108}, 136804 (2012)
\bibitem{SzafranCavity} K. Kolasi\'{n}ski and B. Szafran, Phys. Rev. B. {\bf 88}, 165306 (2013)
\bibitem{Peeters} M. D. Petrovic, F. M. Peeters, A. Chaves and G. A. Farias, J. Phys.: Condens. Matter {\bf 25}, 495301 (2013)
\bibitem{KozikovAB} A. A. Kozikov, R. Steinacher, C. R\"{o}ssler, T. Ihn, K. Ensslin, C. Reichl and W. Wegscheider, New J. Phys. {\bf 16}, 053031 (2014)
\bibitem{Steinacher2015} R. Steinacher, A. A. Kozikov, C. R\"{o}ssler, T. Ihn, K. Ensslin, C. Reichl and W. Wegscheider, New J. Phys. {\bf 17}, 043043 (2015)
\bibitem{SzafranDot} K. Kolasi\'{n}ski and B. Szafran, New J. Phys. {\bf 16}, 053044 (2014)
\bibitem{Aidala} K. E. Aidala, R. E. Parrot, T. Kramer, E. J. Heller, R. M. Westervelt, M. P. Hanson and A. C. Gossard, Nature Physics {\bf 3}, 464 (2007)
\bibitem{Metalidis} G. Metalidis and P. Bruno, Phys. Rev. B {\bf 72}, 235304 (2005)
\bibitem{Cresti} A. Cresti, J. Appl. Phys. {\bf 100}, 053711 (2006)
\bibitem{Metzger} J. J. Metzger, R. Fleischmann and T. Geisel, Phys. Rev. Lett. {\bf 105}, 020601 (2010)
\bibitem{Maryenko} D. Maryenko, F. Ospald, K. v Klitzing, J. H. Smet, J. J. Metzger, R. Fleischmann, T. Geisel, V. Umansky, Phys. Rev. B {\bf 85}, 195329 (2012)
\bibitem{See} A. M. See, I. Pilgrim, B. C. Scannell, R. D. Montgomery, O. Klochan, A. M. Burke, M. Aagesen, P. E. Lindelof, I. Farrer, D. A. Ritchie, R. P. Taylor, A. R. Hamilton, and A. P. Micolich, Phys. Rev. Lett. {\bf 108}, 196807 (2012)
\bibitem{Scannell} B. C. Scannell, I. Pilgrim, A. M. See, R. D. Montgomery, P. K. Morse, M. S. Fairbanks, C. A. Marlow, H. Linke, I. Farrer, D. A. Ritchie, A. R. Hamilton, A. P. Micolich, L. Eaves and R. P. Taylor, Phys. Rev. B {\bf 85}, 195319 (2012)
\bibitem{Micolich} A. P. Micolich {\it et al}, Fortschr. Phys. {\bf 61}, 332 (2013)
\bibitem{Ihn} T. Ihn, ``Electronic Quantum Transport in Mesoscopic Semiconductor Structures", Springer Tracts in Mod. Phys. 192, (2004)
\bibitem{SteinacherPRB} R. Steinacher, A. A. Kozikov, C. R\"{o}ssler, C. Reichl, W. Wegscheider, K. Ensslin and T. Ihn, Phys. Rev. B. {\bf 93}, 085303 (2016)
\bibitem{KozikovNewTechnique} A. A. Kozikov, R. Steinacher, C. R\"{o}ssler, T. Ihn, K. Ensslin, C. Reichl and W. Wegscheider, Nano Lett. {\bf 15}, 7994 (2015)
\bibitem{Heller2} E. J. Heller and S. Shaw, International Journal of Modern Physics B {\bf 17}, 3977 (2003)
\bibitem{Groth} C. W. Groth, M. Wimmer, A. R. Akhmerov and X. Waintal, New J. Phys. {\bf 16}, 063065 (2014)
\bibitem{RoesslerPRL} C. R\"{o}ssler, D. Oehri, O. Zilberberg, G. Blatter, M. Karalic, J. Pijnenburg, A. Hofmann, T. Ihn, K. Ensslin, C. Reichl and W. Wegscheider, Phys. Rev. Lett. {\bf 115}, 166603 (2015)



\end{thebibliography}
\end{document}